\documentclass{article}

\usepackage{arxiv}
\usepackage[utf8]{inputenc}

\usepackage[
    natbib=true,
    style=numeric,
    sorting=none
]{biblatex}
\usepackage{biblatex}
\usepackage[english]{babel}
\usepackage{csquotes}
\usepackage[T1]{fontenc}    
\usepackage{hyperref}
\hypersetup{
    colorlinks=true,
    linkcolor=red,
    filecolor=magenta,      
    urlcolor=blue,
    pdftitle={Overleaf Example},
    pdfpagemode=FullScreen,
    }
\usepackage{authblk}
\usepackage{booktabs}       
\usepackage{amsfonts}       
\usepackage{nicefrac}       
\usepackage{microtype}      
\usepackage{lipsum}		
\usepackage{amsmath}
\usepackage{graphicx}
\usepackage{doi}
\usepackage{graphicx}
\usepackage{dcolumn}
\usepackage{bm}
\usepackage{amssymb}

\title{Force-Free Electromagnetic Configurations in FLRW Geometry}

\author[1,2,\thanks{radhikari@troy.edu}]{Rakshak Adhikari } 
\author[2]{Govind Menon}
\author[1,3,4,5]{Mikhail V. Medvedev}
\affil[1]{Department of Physics, University of Kansas, Lawrence, KS 66045}
\affil[2]{Center for Relativity and Cosmology\\ Troy University, Troy, AL 36082}
\affil[3]{Institute for Advanced Study, School for Natural Sciences, Princeton, NJ 08540}
\affil[4]{Department of Astrophysical Sciences, Princeton University, Princeton, NJ 08544}
\affil[5]{Institute for Theory and Computation, Harvard University, Cambridge, MA 02138}
\affil[5]{Laboratory for Nuclear Science, Massachusetts Institute of Technology, Cambridge, MA 02139}

\renewcommand{\shorttitle}{FFE in the FLRW universe}

\begin{document}
\maketitle
\begin{abstract}
Force-free electrodynamics is the theoretical paradigm used to describe electromagnetic fields in a region where the inertia of plasma is negligible compared to the strength of the electromagnetic field. While these fields are studied extensively around accreting black holes in an attempt to describe energy extraction, force-free fields also routinely appear in the study of cosmological magnetic fields. Despite this, there are no systematic studies of exact force-free fields in an expanding universe. In this paper, we use geometric methods to find a wide variety of force-free solutions in a fixed Friedmann–Lemaître–Robertson–Walker metric background. The method we use can be directly generalized to any arbitrary electrically neutral space-time, and hence, provides a powerful tool to study force-free fields in general.
\end{abstract}


 \section{Introduction}
Magnetic fields are known to permeate the universe from the scales of planets and stars to galaxies and even galaxy clusters \cite{Giovannini:2003yn}. At the scale of galaxies, magnetic fields have a strength of order $10^{-6}$ G and are coherent over kpc scales. Micro-Gauss magnetic fields have also been observed in galaxy clusters \cite{Feretti:2012vk, Bonafede:2010xg}.
Recent observations suggest that intergalactic space may harbor magnetic fields of strength $10^{-16}$ G coherent over Mpc scales. \cite{Neronov:2010gir}.

The currently accepted paradigm explaining the existence of this all-pervading magnetic field states that these magnetic fields originated from the amplification of seed magnetic fields via various astrophysical dynamos. 
While at smaller scales (at the level of planets and stars), these fields require constant rejuvenation to replenish the loss from dissipation, the time scales for dissipation for large-scale magnetic fields may be of the order of the age of the universe. While the amplification of the large-scale magnetic field is attributed to the gravitational collapse of flux-frozen matter during structure formation, the dynamo effect can only amplify a preexisting non-zero "seed" magnetic field. The origin of such seed magnetic field itself is not well understood and is a subject of extensive study, especially over the last two decades. Electroweak phase transition \cite{Kahniashvili:2012uj, Brandenburg:2017neh} and inflationary generation of the seed magnetic fields \cite{Enqvist:2004yy, Jain:2012ga, Subramanian:2009fu} are the most widely studied phenomena for the generation of primordial magnetic fields in the early universe. Other mechanisms for the generation of primordial magnetic fields include Lorentz invariance violation \cite{Bertolami:1998dn}, relativistic positron abundance \cite{Steinmetz:2023nsc}, and non-Gaussian perturbations to the baryon density to name a few \cite{Flitter:2023xql}.

A variety of methods exist for detecting and/or constraining magnetic fields in the universe. At low redshifts, these include the observation of Zeeman splitting \cite{Han2017} and Faraday rotation of linearly polarized radio sources \cite{Xu:2005rb}. At higher redshifts the existence of magnetic fields can be inferred from the effects of primordial magnetic fields on the polarization of the CMB \cite{Planck:2015zrl, Paoletti:2018uic, Pogosian:2018vfr, Sutton:2017jgr}, effects of magnetic pressure on the abundance of light elements during big-bang nucleosynthesis \cite{Kernan:1995bz} as well as (non) detection of inverse-Compton scattered CMB photons from blazar observations \cite{Neronov:2010gir, Essey:2010nd}. Recently, it was shown that hydrodynamical simulations of structure formation in the universe can also constrain the primordial magnetic fields by studying their ability to reproduce in the simulations the scaling relations observed in dwarf galaxies \cite{Sanati:2020oay}.

When the energy stored in the electromagnetic field is much greater than the plasma pressure, the Lorenz force vanishes.  Force-free electrodynamics is the framework that describes such a system. Force-free fields can also be thought of as the low inertia limit of ideal magnetohydrodynamics i.e. the limit when the matter part of the stress-energy tensor can be ignored in favor of a purely electromagnetic stress-energy tensor.

Force-free plasma occurs naturally in various astrophysical contexts. For example, the solar corona is permeated by such strong magnetic fields that it is essentially force-free. The study of force-free fields is also important in the study of relativistic jets. The strong magnetic fields generated around the accreting black hole and neutron stars render these systems force-free and the fields are believed to extract rotational energy from these compact objects to power the highly collimated relativistic jets.

Force-free fields often appear in the study of primordial magnetogenesis \cite{Jedamzik:1999bm, Pusztai2020, Totani:2017zjp, Pavlovic:2018jiz, Maartens:2001zu, Grasso:2000wj, Pogosian:2001np}. Several studies of primordial magnetic fields assume the vanishing of the Lorentz force \cite{Koh:2000qw, Maeda:2008dv, Giovannini:2005jw, Clarkson:2002dd, GarciadeAndrade:2009lqb}. To our knowledge, there has been no significant and systematic attempt to study the types of force-free fields allowed in the Friedmann–Lemaître–Robertson–Walker (FLRW) metric. In this paper we provide a non-exhaustive list of solutions to the equations of force-free electrodynamics (FFE) generated not by solving the arising partial differential equations, but by using the geometric study of the foliations of spacetime as laid out in \cite{Menon:2020ivu}, \cite{Menon:2020npo} and \cite{Menon:2020hdk}.


\section{Equations of Force-Free Electrodynamics}
Maxwell's equation in an arbitrary spacetime is given by
\begin{equation}
 d  F = 0\;, \;\;\;{\rm and}
\;\;\;*d*F = j\;.
\label{inhomMaxform}
\end{equation}
Here $F$ is the Maxwell field tensor, $*$, the Hodge-Star operator, $d$, the exterior derivative on forms, and $j$ denotes the current density dual vector.
Force-free electrodynamics is defined by the constraint $F_{\mu\nu} j^\nu = 0$.
The Maxwell Field tensor $F$ is said to be magnetically dominated whenever $F^2 >0$, $F$ is electrically dominated whenever $F^2 <0$, and finally a force-free electromagnetic field $F$ is null whenever $F^2 =0$. As usual $F^2 =F_{\mu\nu}F^{\mu\nu}$.

The kernel of $F$ is the set of all tangent vectors that annihilate $F$. In force-free electrodynamics, the current density $j$ is always such a vector field. It is well known that the kernel of $F$ forms the tangent space to a two-dimensional submanifold. Starting with this fact, a comprehensive study of geometric features of force-free fields was done in \cite{Menon:2020ivu},  \cite{Menon:2020npo} and\cite{Menon:2020hdk}, and we provide a concise summary in the following sections.

\subsection{The Non-Null Force-Free Field}
Details presented in this section were previously derived in \cite{Menon:2020ivu}. We will first consider the case of a magnetically dominated field. In this case, about any point in spacetime, we can construct an inertial frame field  $(e_0, e_1, e_2, e_3)$ such that
$g(e_\mu, e_\nu) = \eta_{\mu\nu}$, and further
$e_0$ and $e_1$ span the kernel of $F$. Here, $\eta$ is the Minkowski metric. Define vector fields $H$ and $\tilde H$ by

\begin{equation}
 2H= \left[-g(\nabla_{e_0} e_0, e_2)+ g(\nabla_{e_1} e_1, e_2) \right] e_2 + \left[-g(\nabla_{e_0} e_0, e_3)+ g(\nabla_{e_1} e_1, e_3) \right] e_3\;, 
    \label{Hdef}
\end{equation}
and
\begin{equation}
 2\tilde H= \left[-g(\nabla_{e_2} e_2, e_0)- g(\nabla_{e_3} e_3, e_0) \right] e_0 + \left[g(\nabla_{e_2} e_2, e_1)+ g(\nabla_{e_3} e_3, e_1) \right] e_1\;.  
   \label{Htildedef}
\end{equation}

We are guaranteed a magnetically dominated force-free solution if 
\begin{equation}
    d(H^\flat+\tilde H^\flat) = 0\;.
    \label{H+tilH}
\end{equation}

In this case, the magnetically force-free $F$ can now be written as
\begin{equation}
    F= u\;e_2 ^\flat \wedge e_3 ^\flat\;,
    \label{FrameF}
\end{equation}
where
\begin{equation}
 d (\ln u) = 2 (H+\tilde H)^\flat\;. 
 \label{eqforu}
\end{equation}

The $\flat$ maps a tangent vector to its metrically equivalent one-form. The difficulty of finding magnetically dominated force-free fields is now reduced to finding a tetrad frame field that satisfies the above requirements. Hidden in the above statements is the fact that $e_0$ and $e_1$ span smooth two-dimensional submanifolds. These submanifolds are referred to as field sheets.

The electrically dominated field is very similar to the above construction except that it is $e_2$ and $e_3$ that span the kernel of $F$.

\subsection*{The Theory of Null Force-free Fields}

Solutions to null force-free electrodynamics satisfy a different set of criteria stemming from the fact that the kernel of $F$ in this case is spanned by a degenerate distribution. For details see \cite{Menon:2020hdk}.

Here, the kernel of $F$ is spanned by a null geodesic vector field $l$ and a spacelike orthogonal vector field $s$. The tetrad is completed by including another null vector $n$ and a one-form $\alpha$ such that
$$n\cdot l = -1,\;\; 0=l\cdot s =\alpha(l)=n\cdot s= \alpha(n)=\alpha(s),\;\;\alpha^\mu \alpha_\mu = 1 = s^\mu s_\mu\;.$$
In this case, if the equipartition condition for the null mean curvature is satisfied, i.e,
$$g(\nabla_s l,s) = g(\nabla_{\alpha^\sharp} l, \alpha^\sharp)\;,$$
then the force-free null field takes the form
\begin{equation}
F= (u \cdot \kappa)\; \alpha \wedge l^\flat\;.
\label{mixedF}
\end{equation}
Here, $\sharp$ is the inverse map of $\flat$. In an adapted chart $(x^1, x^2, x^3, x^4)$ where field sheets are given by the condition $x^1, x^2 = {\rm const}$,
\begin{equation}
  \kappa=(\alpha_3 \;l^\flat_4-\alpha_4\; l^\flat_3)^{-1}\;,
  \label{kappadef}
\end{equation}
where 
\begin{equation}
  \left(
           \begin{array}{c}
             \alpha \\
             l^\flat \\
           \end{array}
         \right)=\left(
    \begin{array}{cc}
      \alpha_3 & \alpha_4 \\
      l^\flat_3 & l^\flat_4  \\
    \end{array}
  \right)\left(
           \begin{array}{c}
             dx^3 \\
             dx^4 \\
           \end{array}
         \right)\;. 
         \label{chart2frame}
\end{equation}
Here $u = u(x^3, x^4)$ is any smooth function of $x^3, x^4$.


\section*{Force-free Solutions}

The conformal time coordinate chart often simplifies the calculations in FLRW spacetimes. In this coordinate system, the metric takes the following form:
\begin{equation}
    ds^2=a(\eta)^2\bigg[ -d\eta^2 +\frac{dr^2}{1-K\:r^2}+ r^2  \left(d\theta^{2} + \sin^{2}\theta\; d\varphi^2 \right)\bigg]\;.
\end{equation}
Since we will rely primarily on the tetrad formalism described above, we begin by listing a set of canonical orthonormal tetrad for the metric given above:
\begin{subequations}
\begin{align}
  e_0 &=\frac{1}{a(\eta)}\partial_\eta,  \\
  e_1 &=\frac{ \sqrt{1-Kr^2}}{a(\eta)}\:\partial_r,\\
 e_2 &= \frac{1}{a(\eta)\:r}\:\partial_\theta\;,\\
 e_3 &=\frac{}{a(\eta)\:r\:\sin{\theta}}\:\partial_\varphi\;.
\end{align}
\label{F2L2}
\end{subequations}

\subsection*{Solution I}
Consider a Lorentz transformation of the canonical tetrads given by
\begin{align}
\begin{pmatrix}
\bar e_0\\
\bar e_1\\
\bar e_2\\
\bar e_3\\
\end{pmatrix}=
\begin{pmatrix}
\alpha \sin(\theta)  f_2 & 0 & r\:f_1   f_2 & r \: f  f_2 \\
0 & 1 & 0 & 0 \\
\sqrt{\beta} \: r  f_2 & 0 & \dfrac{f_1  \alpha \sin(\theta) \: f_2}{\sqrt{\beta}} & \dfrac{\alpha \sin(\theta) \: f \: f_2}{\sqrt{\beta}} \\
0 & 0 & \dfrac{f}{\sqrt{\beta}} & -\dfrac{f_1}{\sqrt{\beta}}
\end{pmatrix}
\begin{pmatrix} e_0\\
e_1 \\
e_2\\
e_3\\
\end{pmatrix},\;
\end{align}
where $f$ is any function of $r$, $\alpha$ and $\beta$ are real constants, and $$f_1=\sqrt{\beta-f^2}\;\;\; {\rm and} \;\;\; f_2=\dfrac{1}{\sqrt{\alpha^2\sin^2{\theta}-\beta\:r^2}}.$$
Then the pair of vector fields $(\bar e_2,\bar e_3)$ are involutive and further
\begin{equation}
    2(H+\Tilde{H})=\frac{2\dot a(\eta)}{a(\eta)} \:d\eta +\frac{2\:\alpha^2\:\sin^2{\theta}-\beta\:r^2}{r(\beta\:r^2-\alpha^2\sin^2{\theta})}\:dr+\frac{\beta^2\:r^2\:\cot{\theta}}{\alpha^2\:\sin^2{\theta}-\beta\:r^2}d\theta\;.
\end{equation}
It is easy to verify that $d(H+\Tilde{H})=0.$ Then from eqs.(\ref{H+tilH}), (\ref{FrameF}) and (\ref{eqforu}),
\begin{equation}
    u=\frac{u_0\sqrt{\alpha^2\:\sin^2{\theta}-\beta\:r^2}}{a(\eta)^2\:r^2\:\sin{\theta}},
\end{equation}
and our electrically dominated solution is given by

\begin{align}
    F_1&= u\:e_0^\flat \wedge e_1^\flat\\ \nonumber
&=\frac{u_0\:\alpha}{ r^2 \sqrt{1-K r^2}} d\eta \wedge dr+\frac{u_0\sqrt{\beta-f^2}}{\sin{\theta}\sqrt{1-Kr^2}} dr \wedge d\theta+\frac{u_0\:f}{\sqrt{1-Kr^2}} dr \wedge d\varphi\;.
\end{align}
Here, the current density is given by
\begin{align}
    j_1=\frac{\sqrt{1-K\:r^2}f'\:}{a(\eta)^4\:r^2\:\sin{\theta}} \Bigg[\frac{f}{\sqrt{\beta-f^2}}\partial_\theta-\csc{\theta}\partial_\varphi\Bigg].
\end{align}
The Lorentz invariant quantity $F^2$ in this case is given by
\begin{equation}
    F_1^2=-\frac{2\:u_0^2\:\left(\alpha^2\:\sin^2{\theta}-\beta r^2\right)}{a(\eta)^4\:r^4 \: \sin^2{\theta}}.
\end{equation}
From the above equation, we see that the solution is not well defined when $\sin \theta =0$. For completeness, we also record the magnitude of the square of the current density vector: 
\begin{equation}
    j^2 _1=\frac{u_0^2\:\left(K \,r^{2}-1\right) {f'}^{2} \beta}{a(\eta)^6\:r^2\:\sin^2{\theta}(f^2-\beta)}.
\end{equation}

This solution holds some intriguing features that deserve closer inspection. First, we note that the Lorentz transformation that generates the solution is not valid when
$$\chi \equiv \alpha^2\sin^2{\theta}-\beta r^2 \leq 0\;.$$ Nonetheless, $F_1$ does not depend on $\chi$. This means that the solution $F_1$ is defined for all values of $\chi$ (except when $\sin\theta =0$ for an entirely different reason). As it turns out, examining $F_1 ^2$ tells us that our electrically dominated solution smoothly transitions to a null solution when $\chi=0$ and further a magnetically dominated solution when $\chi<0$.
The tetrad formalism is unsuitable to handle such a transition. So we shift to a different formalism using a foliation-adapted chart which was developed in \cite{Menon:2020npo}. About any point in spacetime, there exists a coordinate chart $(x^1, \dots, x^4)$ that is adapted to field sheets meaning that the tangent space of the submanifolds defined by constant values of $x^1$ and $x^2$ contain the kernel of $F$.
Then $F$ takes the form
\begin{equation}
F= u(x^3, x^4) \;dx^3 \wedge dx^4\;.
\label{finalformF}
\end{equation}
Here, let
$$M^r\equiv g^{r 3} \;g^{3 4}- g^{3 3} \;g^{r 4}\;, \;\;\;{\rm and}\;\;\;N^r \equiv g^{r 3} \;g^{4 4}- g^{3 4} \;g^{r 4}\;.$$
\vskip0.2in \noindent
Then as it was shown in \cite{Menon:2020npo}, the equations of FFE in this coordinate system are given by
\begin{equation}
M^4\;\frac{\partial}{\partial{x^4}} \ln |u| =-\nabla_r\; M^r
\label{FF v3_a}
\end{equation}
and
\begin{equation}
N^3\;\frac{\partial}{\partial{x^3}} \ln |u| =-\nabla_r\; N^r\;.
\label{FF v3_b}
\end{equation}
\vskip0.2in \noindent
To obtain a foliation-adapted chart for the case of this solution, define vector fields

$$X_1 = \sin\theta \; \partial_\theta  -\frac{\sqrt{\beta -f^2} }{f}\, \partial_\varphi\;,\;\;\;\;X_2 = \partial_\eta  -\frac{\alpha }{r^2 f}\, \partial_\varphi\;,$$
$$X_3 =\partial_\eta +\partial_r +\left[\left(-\beta\ln(\csc \theta +\cot\theta)\frac{f' }{f^2 \sqrt{\beta - f^2}}\right)-\alpha \eta \left(\frac{2}{r^3 f}+\frac{f'}{r^2 f^2}\right)\right]\partial_\varphi\;,$$
and finally 
$$X_4 = \partial_\varphi\;.$$
\vskip0.2in \noindent
It is easily verified that all of the vector fields $\{X_i\}$ defined above commute with each other. Therefore there exist coordinate functions $\{x^i\}$ such that $X_i =\partial_{x^i}$ for each $i$. The dual bases $\{dx^i\}$ are such that

$$dx^1 = \csc\theta d\theta\;,\;\;\;\; dx^2 =d\eta - dr, \;\;\;\;dx^3 = dr\;,$$
and

$$dx^4 =  -\frac{\alpha }{r^2 f} d\eta +\left[\frac{\alpha}{r^2 f}+\frac{2 \alpha \eta}{r^3 f}+\frac{\alpha \eta f'}{r^2 f^2}+\frac{\beta \ln(\csc \theta +\cot\theta)f'}{f^2 \sqrt{\beta -f^2}}\right]dr +\frac{\sqrt{\beta -f^2}}{\sin \theta f} d\theta +d\varphi\;.$$
Then
$$dx^i(\partial_{x^j}) = \delta^i _j$$
as required.
This will help us compute the determinant of the metric in the adapted basis and also the quantities $M^r$ and $N^r$ as defined above.  The relevant quantities in eqs. (\ref{FF v3_a}) and (\ref{FF v3_b}) are given by

$$\sqrt{-g} = \frac{ a^4(\eta) r^2 \sin^2 \theta}{\sqrt{1-Kr^2}}\;,$$

$$M^1 =  \frac{-\sqrt{\beta -f^2} (1-Kr^2)} { a^4(\eta) r^2 \sin^2 \theta f}\;, \;\;\;M^2 = \frac{-\alpha(1-Kr^2)} { a^4(\eta) r^2  f}, \;\;\;M^3 =0\;,$$

$$M^4 =  \frac{ (1-Kr^2) \chi} { a^4(\eta) r^4 \sin^2 \theta f^2}  =-N^3\;,$$

$$N^1 =  \frac{- (1-Kr^2) } { a^4(\eta) r^5 \sin^2 \theta f^3} \left[\Big(f' \eta r + (r + 2\eta)\:f\Big) \alpha\sqrt{\beta -f^2} +\beta \ln(\csc \theta +\cot\theta)f' r^3\right] \;,$$
and finally

$$N^2 = \frac{  (Kr^2-1) } { a^4(\eta) r^5  f^3 \sin^2 \theta } \left[\left(\alpha \eta  + \beta r^2\frac{\ln(\csc \theta +\cot\theta)}{\sqrt{\beta -f^2}} \right) \sin^2\theta r \alpha f^\prime - f\left(\beta r^3 + 2 \alpha^2 \eta \sin^2\theta\right)\right] \;.$$

Noting that $u = u(x^3, x^4 = \varphi)$, eq.(\ref{FF v3_a}) becomes
$$M^4\;\frac{\partial}{\partial{\varphi}} \ln |u| = 0\;.$$
I.e., $u_{, \varphi} =0$ and eq.(\ref{FF v3_b}) gives

\begin{equation}
   \chi \frac{ (1-Kr^2) } { a^4(\eta) r^4 \sin^2 \theta f^2} \frac{d}{d{x^3}} \ln |u| = \chi \left[\frac{ f' (1-Kr^2) +K r f } { a^4(\eta) r^4 \sin^2 \theta f^3}\right]\;.
    \label{FF v3_c}
\end{equation}

Notice how the factor $\chi$ cancels out in the above equation, and hence we can smoothly transition from an electrically dominated solution to a magnetically dominated one. \footnote{Henceforth such solutions will be referred to as type-changing solutions.}

The above equation is satisfied by setting $$u = \frac{u_0\:f}{\sqrt{1-Kr^2}}\;.$$
Therefore, in the adapted chart
$$F_1 =\frac{u_0\:f}{\sqrt{1-Kr^2}}\; dx^3 \wedge dx^4\;.$$

Using the method of tetrads and searching for Lorentz transformations that satisfy eqs.(\ref{H+tilH}), (\ref{FrameF}) and (\ref{eqforu}), we have been able to generate several non-trivial solutions in FLRW spacetimes. In the remainder of this section, we simply list the solutions without referring to the generating Lorentz transformation.

\subsection*{Solution II}
 Using a time-dependent Lorentz transformation we obtain the following electrically dominated solution
\begin{equation}
    F_2=\frac{\alpha}{r^2\sqrt{1-K\:r^2}}d\eta\wedge dr+\frac{\sqrt{\beta-f^2}}{\sin{\theta}} d\eta \wedge d\theta\:+f\: d\eta \wedge d\varphi.
\end{equation}
Here $f=f(\eta)$, and, $\alpha$ and $\beta$ are real constants. The current density is then given by
\begin{equation}
    j_2=\frac{f'}{a^4\:  r^2\:\sin{\theta}} \left(\frac{-f}{\sqrt{\beta-f^2}}\partial_\theta+\frac{\partial_\varphi}{\sin{\theta}}\right).
\end{equation}
The Lorentz invariant scalars in this case are
\begin{equation}
    F_2^2=-\frac{2\alpha^2\:\sin^2{\theta}+2\beta\:r^2}{a^4\:r^4\sin^2{\theta}},
\end{equation}
and 
\begin{equation}
    j_2^2=\frac{{f'}^{2}\:\beta}{a^6\:\sin^2{\theta}(\beta-f^2)}.
\end{equation}
As in the previous case, this solution is not well defined when $\sin \theta =0$.
\subsection*{Solution III}
For, $K=1$, and  $f=f(\theta)$, and a real constant $\alpha$, we obtain another type changing solution of the form
\begin{align}
    F_3=&\frac{(\sqrt{K}r+\sqrt{Kr^2-1})^\alpha}{\sin{\theta}} \bigg[\left(f\:\sin{(\alpha\sqrt{K}\:\eta)}-\sqrt{\beta-f^2}\cos{(\alpha\sqrt{K}\:\eta)} \right)d\eta \wedge d\theta \nonumber \\
    &-\frac{1}{\sqrt{K\:r^2-1}}\left(f \:\cos{(\alpha\sqrt{K}\:\eta)}+\sin{(\alpha\sqrt{K}\:\eta)}\sqrt{\beta-f^2}\right)\: dr\wedge d\theta \bigg] \nonumber \\
    &+C\:\sin{\theta}\; d\theta \wedge d\varphi .
\end{align}
Here, the current density is given by
\begin{align}
    j_3=&\frac{\left(\sqrt{K}\, r +\sqrt{K \,r^{2}-1}\right)^{\alpha} f}{a \! \left(\eta \right)^{4} \sin \! \left(\theta \right) r^{2} \sqrt{\beta-f^{2}}}\bigg[-\left(f \cos \! \left(\alpha  \sqrt{K}\, t \right)-\sin \! \left(\alpha  \sqrt{K}\, t \right) \sqrt{\beta-f^{2}}\right)\:\partial_\eta \nonumber \\ 
    &+\left(-f \sin \! \left(\alpha  \sqrt{K}\, t \right)+\sqrt{\beta-f^{2}}\, \cos \! \left(\alpha  \sqrt{K}\, t \right)\right) \sqrt{K \,r^{2}-1}\:\partial_r\bigg].
\end{align}

In this case

\begin{equation}
    F_3^2=-\frac{2 \left(\sqrt{K}\, r +\sqrt{K \,r^{2}-1}\right)^{2 \alpha} \beta  \,r^{2}-2C^2 \sin^2 \theta }{a \! \left(\eta \right)^{4} r^{4} \sin^2{\theta}}\;.
\end{equation}

Once again this solution is not well defined when $\sin \theta =0$. Further, when $K=0, -1$, the solution above does indeed satisfy the force-free Maxwell's equation. However, the coefficient terms become complex, making the solution physically irrelevant. 
\subsection*{Solution IV}
We now present the following electrically dominated force-free field
\begin{equation}
    F_4=f_1\:d\eta \wedge d\theta+f_2\: d\eta \wedge d\varphi +f_3\:d\theta \wedge d\varphi,
\end{equation}
where,
\begin{equation}
    f_1=\frac{{\mathrm e}^{-\frac{\eta}{\omega_{2}}}}{\sin{\theta}} \left(\left(2 \beta  \alpha  \cos^2{\theta}+\left(2 \alpha  k_{1}+2 \beta  k_{2}\right) \cos{\theta}+k_{3}\right) {\mathrm e}^{\frac{2 \eta}{\omega_{2}}}-\left(\cos{\theta} \beta +k_{1}\right)^{2} {\mathrm e}^{\frac{4 \eta}{\omega_{2}}}-\left(\cos{\theta} \alpha +k_{2}\right)^{2}
\right)^{1/2},
\end{equation}

\begin{equation}
    f_2=\left(\cos{\theta} \alpha +k_{2}\right) {\mathrm e}^{-\frac{\eta}{\omega_{2}}}+{\mathrm e}^{\frac{\eta}{\omega_{2}}} \left(\cos{\theta} \beta +k_{1}\right),
\end{equation}
and
\begin{equation}
    f_3=-\omega_{2} \sin{\theta} \left({\mathrm e}^{\frac{\eta}{\omega_{2}}} \beta -\alpha  \,{\mathrm e}^{-\frac{\eta}{\omega_{2}}}\right).
\end{equation}
Here $k_1,k_2,\omega_1$ and $\omega_2$ are real constants
The current density is given by

\begin{align}
    j_4= &\frac{1}{a(\eta)^4\:r^2} \Bigg[-(f_1\:\cot{\theta}
+\partial_\theta f_1) \: \partial_\eta+\:(\partial_\eta f_1) \partial_\theta+\: \frac{\csc{\theta}^2}{r^2}\left(   \partial_\eta f_2-\partial_\theta f_3 +f_3 \:\cot{\theta}\right)\:\partial_\varphi  \Bigg].
\end{align}

Here
\begin{align}
    F_4^2=-\frac{2}{a(\eta)^4\:r^4\:\sin^2{\theta}}\Bigg[ \sin^2{\theta}\:r^2\:f_1^2+r^2f_2^2-f_3^2\Bigg],
\end{align}
and 
\begin{align}
    j_4^2=\frac{1}{a(\eta)^6\:r^6}\Bigg[ \csc{\theta}\:(r^2\:\partial_\eta f_2-\partial_\theta \: f_3+f_3\:\cot{\theta})^2+(\partial_\eta \: f_1)^2 -r^2\:(f_1\:\cot{\theta}+\partial_\theta\:f_1)^2\Bigg].
\end{align}

This solution is not valid during early and late times in addition to the usual pathology when $\sin \theta =0$.
\subsection*{Solution V and VI}
We continue with our presentation by illustrating a few null solutions in FLRW spacetimes. The tetrad formalism for generating null solutions is substantially different from that of non-null solutions. As described in the aforementioned section, we begin with a null pre-geodesic congruence defined by

\begin{equation}
    l=\frac{\csc{\theta}}{r\:a(\eta)^3 }\: \left(\partial_\eta-\sqrt{1-K\:r^2}\:\partial_r\right).
\end{equation}

Here, $l$ is a null vector satisfying $\nabla_l\:l \propto l$. A simple calculation shows that
\begin{equation}
    \nabla_l\:l=\frac{\dot a(\eta)\:r+a(\eta)\sqrt{1-k\:r^2}}{r^2\:\sin{\theta}\:a(\eta)^4} \: l.
\end{equation}

The relevant null tetrad $(s,l,\alpha^\sharp,n)$ can be constructed by defining the following vector fields:

\begin{equation}
    n=\frac{a(\eta)\:r\:\sin{\theta}}{2}\left(\partial_\eta+\sqrt{1-K\:r^2}\:\partial_r\right),
\end{equation}

\begin{equation}
    \alpha^\sharp=\frac{1}{r\:a(\eta)}\:\partial_\theta
\end{equation}

\begin{equation}
    s=\frac{\csc{\theta}}{r\:a(\eta)}\:\partial_\varphi
\end{equation}
Here, since
\begin{equation}
    g(\nabla_s l,s)=g(\nabla_{\alpha^\sharp}l,\alpha^\sharp)=\frac{\dot a(\eta)\:r-a(\eta)\sqrt{a-Kr^2}}{a(\eta)^4\:r^2\:\sin{\theta}}\;,
\end{equation}
we see that the equipartition condition for the null mean curvature is satisfied. Additionally, since $l,s$ forms an involutive distribution, we are guaranteed the existence of a null force-free field solution with a two-parameter prefactor. To isolate the prefactor, we construct a foliation-adapted chart with commuting vector fields defined as follows
$$X_1=\partial_\eta+\sqrt{1-Kr^2},\;\;\;X_2=\partial_\varphi,\;\;\; X_3=\partial_\theta, \;\;\;{\rm and }\;\;\; X_4=l.$$ Our adapted chart is then defined by the following coordinate one-forms

$$dx^1=d\eta-\frac{dr}{\sqrt{1-Kr^2}},\;\;\;dx^2=d\varphi,\;\;\;dx^3=d\theta \;\;\;{\rm and}\;\;\;dx^4=d\eta+\frac{dr}{\sqrt{1-Kr^2}}.$$

The null and force-free field is then given by
\begin{align}
    F_5&= (u\cdot \kappa)\: \alpha \wedge l^\flat
    = u(x^3,x^4)\:\left(d\eta\:\wedge d\theta+\frac{1}{\sqrt{1-Kr^2}}\:dr \wedge d\theta\right)\;.
\end{align}

Here, $\kappa=-\csc{\theta}$ and is given by \ref{kappadef}, and the current density is then given by

\begin{equation}
    j_5=\frac{\partial_\theta\:f+f\:\cot{\theta}}{a(\eta)^4\:r^2}\left(-\partial_\eta+\sqrt{1-Kr^2}\:\partial_r\right).
\end{equation}
As required, $F_5^2=0$, and since the current density is along $l$ we have that $j_5^2=0$. It turns out that this null geodesic congruence also satisfies the uniform equipartition condition described in \cite{Menon:2020hdk}:
\begin{equation}
g(\nabla_s\:l,\alpha^{\sharp})+g(\nabla_{\alpha^\sharp}\: l,s)=0.
\end{equation}
This allows for the possibility of a generalized null solution, which in this case is given by
\begin{equation}
    F_6= \partial_\theta f(\bar t,\theta,\varphi)\:\left(d\eta \wedge d\theta+\dfrac{dr\wedge d\theta}{\sqrt{1-Kr^2}}\right)+\partial_\varphi f(\bar t,\theta,\varphi)\:\left(d\eta \wedge d\phi+\dfrac{dr\wedge d\phi}{\sqrt{1-Kr^2}}\right)\;.
\end{equation}
The $t$ in the above expression is the usual cosmic time given by the relation
\begin{align} \label{tbar}
    &\bar t=\eta+\int \frac{1}{\sqrt{1-K\:r^2}} dr\;.
\end{align}
The current density in this case is given by

\begin{equation}
j_6=\dfrac{\csc{\theta}\partial_\phi^2\:f+\partial_\theta^2f+\cot{\theta}\partial_\theta f}{a(\eta)^4\:r^2}\left(\partial_\eta-\sqrt{1-K\:r^2}\partial_r\right)\;.
\end{equation}

For clever choices of $f$, the solution above is valid when $r \neq 0$.
\subsection*{Solution VII}

For $f=f(\bar t,\theta)$, where $t$ is the cosmic time function defined above, we have the following non-null force-free solution with null current
\begin{equation}
    F_7=f\:\sin{\theta}\left(d\eta \wedge d\theta+\dfrac{1}{\sqrt{1-K\:r^2}} \:dr \wedge d\theta \right)+\alpha\:\sin{\theta}\:d\theta \wedge d\phi\;.
\end{equation}
Here, as usual, $\alpha$ is a real constant. This solution turns out to be a non-null generalization of our previous null solution $F_5$. The current density in this case is given by

\begin{equation}
    j_7=-\dfrac{2\:\cos{\theta}\:f+\sin{\theta}\:\partial_\theta f}{a(\eta)^4\:r^2}\left(\partial_\eta+\sqrt{1-Kr^2}\:\partial_r\right)\;.
\end{equation}

We also have that
\begin{equation}
    F^\sharp_7=-\dfrac{f\:\sin{\theta}}{a(\eta)^4\:r^2}\:\partial_\eta \wedge \partial_\theta+\dfrac{f\:\sin{\theta}\sqrt{1-Kr^2}}{a(\eta)^4\:r^2}\: \partial_r\wedge\partial_\theta+\dfrac{\alpha}{a(\eta)^4\:r^4\sin{\theta}}\:\partial_\theta \wedge\partial_\phi.
\end{equation}
While it appears that the last term in the right-hand side of the above equation is undefined when $\sin \theta =0$, the contraction of the Faraday tensor with itself does not suffer from the same pathology i.e.

\begin{equation}
    F_7^2=\dfrac{2\alpha^2}{a(\eta)^4\:r^4}.
\end{equation}

This is an indication that the solution may just have a coordinate singularity along $\sin \theta =0$. We demonstrate this fact by explicitly transforming it into a Cartesian coordinate system for the case when $K=0$. In the usual $(\eta, x, y, z)$ coordinate system, where we have just transformed the spatial spherical coordinates to Cartesian coordinates,
\begin{align}
    F_7=&\dfrac{1}{r^3}\:\big(zxf\:d\eta \wedge dx+zyf\:d\eta \wedge dy-(x^2+y^2)\:f\:d\eta \wedge dz \nonumber\\&\alpha z \: dx \wedge dy-(\alpha y+rxf)\:dx \wedge dz+(x\alpha-ryf)\: dy \wedge dz \big)\;.
\end{align}
 The above expression is well-defined along the z-axis.
\section*{Solution VIII}
The FLRW metric is further simplified when we set $K=0$, and in the cartesian coordinates described above, we are able to find two new non-null solutions.
First, for constants $c_1, c_2, c_3, c_4$, we have the following force-free field 
\begin{equation}
    F_8= \frac{\sqrt{(-c_1^2+c_2^2)f^2+c_2c_4^2}}{c_2}\:d\eta \wedge dx+\frac{c_1\:f}{c_2} \: d\eta \wedge dz+f\: dx \wedge dz,
\end{equation}
where,
\begin{equation}
    f=f(c_1\:\eta+c_2\:x+c_3).
\end{equation}
The current density is given by
\begin{equation}
    j_8=\frac{(c_1^2-c_2^2)(\partial_\eta f)}{a(\eta)^4\:\sqrt{(c_2^2-c_1^2)f^2+c_4c_2^2}} \bigg[\frac{f}{c_1}\partial_\eta -\frac{f}{c_2}\:\partial_x+\frac{\sqrt{(c_2^2-c_1^2)f^2+c_4c_2^2}}{c_1c_2}\:\partial_z\bigg].
\end{equation}
The Lorentz scalars of the theory are given by

\begin{equation}
    F_8^2=-\frac{2c_4}{a(\eta)^4},
\end{equation}
 and
\begin{equation}
    j_8^2= -\frac{c_4\:(c_1^2-c_2^2)^2\:(\partial_\eta f)^2}{c_1^2\:a(\eta)^6\:((c_1^2-c_2^2)f+c_4c_2^2)}.
\end{equation}
\subsection*{Solution IX}
As in the Cartesian case of the above example, we now provide a secondary non-null solution given by
\begin{equation}
    F_{9}=\frac{\sqrt{c_4\:c_2^2-(c_2^2+c_1^2)f^2}}{c_2} \:dx\wedge dy+\frac{c_1\:f}{c_2}\: dx\wedge dz+f\:dy \wedge dz\;.
\end{equation}
Here $f=f(c_1\:x+c_2\:y+c_3)$. The current density is given by
\begin{equation}
    j_{9}=\frac{(c_1^2+c_2^2)\:(\partial_x\:f)}{a(\eta)^4\:\sqrt{(c_1^2+c_2^2)f^2-c_4c_2^2}}\:\bigg[-\frac{f}{c_1}\:\partial_x+\frac{f}{c_2}\:\partial_y-\frac{\sqrt{c_4\:c_2^2-(c_2^2+c_1^2)f^2}}{c_1c_2}\:\partial_z\bigg].
\end{equation}

And we have
\begin{equation}
    F_{9}^2=\frac{2\:c_4}{a(\eta)^4},
\end{equation}
and,
\begin{equation}
    j_{9}^2=- \frac{(\partial_x f)^2(c_1^2+c_2^2)^2\:c_4}{c_1^2\:a(\eta)^6\:((c_1^2+c_2^2)f^2-c_4\:c_2^2)}.
\end{equation}

A simple spatial rotation in the $x,y$ plane can simplify the expression for $F_9$. Consider the transformation given by

\begin{equation}
    \begin{bmatrix}
        x'\\y'
    \end{bmatrix}=\frac{1}{\sqrt{c_1^2+c_2^2}}\begin{bmatrix}c_2&-c_1\\c_1&c_2
  \end{bmatrix} \begin{bmatrix}
        x\\y
    \end{bmatrix}.
\end{equation}

After the transformation $f=f(c_3+y'\sqrt{c_1^2+c_2^2})$. The constants can be absorbed by redefining $f$ as $f(c_3'+y')$. The solution $F_9$ then takes the form

\begin{align}
    F_{9}'&=\dfrac{\sqrt{c_4c_2^2-(c_1^2+c_2^2)f'^2}}{c_2}\:dx'\wedge dy'+\dfrac{f'\sqrt{c_1^2+c_2^2}}{c_2}\:dy'\wedge dz. 
\end{align}
By redefining the constants, we can rewrite the solution as

\begin{equation}
    F_{9}'=\sqrt{c^2-f^2}\: dx'\wedge dy'+f \:dy'\wedge dz\;.
\end{equation}

The co-moving observer with four-velocity $v^\mu=a(\eta)^{-1}\partial_\eta$ does not see an electric field,   in $F_{9}'$, while the magnetic field is given by \footnote{ In the $3+1$ formalism, the electric and magnetic fields are given by the expression $E_\mu = v^\nu F_{\mu\nu}$ and $B_\mu=v^\nu\ast  F_{\mu\nu}$.}
$$B^x=-\dfrac{f}{a(t)^3}\;\;\;{\rm and}\;\;\;B^z=-\dfrac{\sqrt{c^2-f^2}}{a(t)^3}\;.$$
This solution describes slabs of uniform magnetic field that lie in the $xz$ plane and the field orientation varies in the perpendicular ($y$) direction. As the magnetic field strength is constant throughout space but the field direction changes, this field configuration describes the magnetohydrodynamic `tangential discontinuity'. Indeed, the structure is force-free since the field lines have no tension force (there is no field line bending) and there are no magnetic pressure gradients. Now, a posteriori, it seems easy to understand that the tangential discontinuity remains a force-free solution in the FLRW spacetime since uniform expansion doesn’t change the field topology but instead simply rescales the field strength. 

\section{Conclusion}
In this paper, we have demonstrated the geometric methods of foliations to generate both null and non-null force-free electromagnetic fields for the FLRW spacetime. The equations of force-free electrodynamics are in general complex nonlinear partial differential equations, and exact solutions are very difficult to come by. Even in the extensively studied Schwarzschild and Kerr spacetimes, where force-free electrodynamics is expected to describe the tenuous plasma around accreting black holes, very few exact solutions are known. 
In \cite{Adhikari:2023hhk} we presented several exact solutions to FFE in Kerr spacetime that were generated by the study of foliations. Using the same methods a wide range of exact force-free fields allowed by the FLRW geometry were found. The pathology along the z-axis seems to be a common feature of most of the force-free solutions in both the spacetimes. We have presented, to the best of our knowledge, the first force-free field that transitions from electrically dominated to null, and then to a magnetically dominated regime. We have done so using a chart adapted to the foliation generating the force-free field as such solutions are not handled by the tetrad formalism that describes the different geometrical properties of the null and non-null solutions.

\section{Acknowledgements}
Some of the computations in this paper were performed by using Maple™. Maple is a trademark of Waterloo Maple Inc.

\printbibliography
\end{document}